# ROSAT observations of PSR 2334+61 in the supernova remnant G114.3+0.3

**W. Becker[1], K.T.S. Brazier[2], and J. Trümper[1]**

[1] Max-Planck-Institut für extraterrestrische Physik, 85740 Garching bei München, Germany
[2] Institute for Astronomy, University of Edinburgh, Royal Observatory, Blackford Hill, Edinburgh, EH9 3HJ, U.K.

**Abstract.** We report on the observation of PSR 2334+61 in the soft X-ray domain with ROSAT. Assuming a power-law spectrum $dN/dE \propto E^{-\alpha}$ with photon-index $\alpha = 2$ we derive an X-ray flux of $f_x = (7.1\pm0.2)\times10^{-14}$erg s$^{-1}$cm$^{-2}$ within the 0.1–2.4 keV energy range. An upper limit for the neutron star's surface temperature is put at $T_s^\infty \sim 1.2\times10^6$ K for a neutron star with a medium stiff equation of state (FP-model with $M = 1.4$ M$_\odot$, $R = 10.85$ km). Slightly different values for $T_s^\infty$ are computed for the various neutron star models available in the literature, reflecting differences in the equation of state. No soft X-ray emission is detected from the supernova remnant G114.3+0.3 associated with PSR 2334+61.

**Key words:** Pulsars: individual (PSR 2334+61) – X-rays: general – Stars: neutron – Supernovae and supernova remnants: individual (G114.3+0.3)

## 1. Introduction

Nearly 700 neutron stars are currently known as rotation-powered pulsars (Taylor et al 1995). In the course of the ROSAT mission, about 10% of them have been observed in detailed pointed observations, leading to the detection of pulsed X-ray emission from 10 of them, while 9 more pulsars are identified only by positional coincidence with a radio pulsar (see Becker 1995 and Ögelman 1995 for references). The collective X-ray emission properties of the ROSAT detected pulsars suggest that their X-ray emission characteristics depend mainly on the object's age. Young pulsars with a characteristic age of about $1-2\times10^3$ years emit sharp X-ray pulses with a pulsed fraction of up to 100%. Their X-ray emission is thought to be connected with the acceleration of electrons/positrons in the pulsar magnetosphere and they have spectra well described by a power-law $dN/dE \propto E^{-\alpha}$ with photon index $\alpha \approx 2$. No thermal X-ray emission from the surface of these young pulsars could be detected. The observation of thermal X-rays is complicated by the fact that the youngest and hottest neutron stars are also those with the strongest magnetospheric emission which buries the thermal component (Becker & Aschenbach 1995). Different from these Crab-like pulsars are pulsars having an age between $10^4$-$10^5$ years like PSR 2334+61. They show smoother lightcurves and were found to resemble more the X-ray emission properties of the Vela-pulsar; i.e. strong steady emission from a pulsar powered synchrotron nebula combined with a small pulsed contribution from magnetospheric or thermal emission (Becker, Barzier & Trümper 1995, Ögelman, Finley & Zimmermann 1993). For the Vela pulsar, the spectrum of the pulsed X-ray emission is of thermal origin and contributes only below $\sim 1.2$ keV (Ögelman 1995). The emission from the synchrotron nebula dominates the Vela pulsar spectrum above $\sim 0.5$ keV and follows a power-law with photon-index $\alpha \approx 2$.

In this letter, we report on the ROSAT observation of PSR 2334+61. Soft X-ray emission from this pulsar has been detected recently by Becker, Trümper & Ögelman (1993a). The $\sim 41,000$-year old pulsar belongs to the oldest pulsars that are associated with a supernova remnant. Its spin period of 495 ms is considerably larger than the periods of other SNR-associated pulsars, but is counterbalanced by a large rate of change $\dot{P} = 191\times10^{-15}$ s s$^{-1}$ that implies an unusually strong magnetic field of $\sim 10^{13}$ G and a spin-down power of $\dot{E} = 6.3\times10^{34}$ erg s$^{-1}$. The association of PSR 2334+61 with the supernova remnant G114.3+0.3 was recognized only recently by a cross-correlation of the Princeton-Pulsar-Catalogue with SIMBAD (Kulkarni et al. 1993) and the Effelsberg 11 cm survey (Fürst, Reich & Seiradakis 1993). The pulsar distance of $\sim 2.5$ kpc as determined from the pulsar's dispersion measure using the electron density model of Taylor & Cordes (1993) is found to be consistent with the distance of $\sim 2.5-3$ kpc estimated for the supernova remnant G114.3+0.3 (Reich & Braunsfurth 1981).

*Send offprint requests to*: web@mpe-garching.mpg.de

PSR 2334+61 was observed with the ROSAT PSPC over the period 1992 July 27 to August 4, for a total exposure time of 8350 s. A weak point source was detected within 10 arcsec of the pulsar's radio position. The net count rate, after corrections for background, vignetting and deadtime, was determined to be $(0.0018 \pm 0.0005)$ count/s. No soft X-ray emission is detected from the supernova remnant.

The small number of detected X-ray counts from the pulsar unfortunately precludes any detailed spectral analysis of the source or testing for modulation at the 495 ms spin period. Some limited analysis is possible, however, if we assume that PSR 2334+61 is like other ROSAT pulsars of similar age, whose emission spectra can be of thermal or magnetospheric origin (Becker 1995, Ögelman 1995). Under the assumption that the pulsar's X-ray emission is dominated by the emission of a pulsar-powered synchrotron nebula which has a spectrum similar to that of the Vela-nebula, i.e. follows a power-law $dN/dE \propto E^{-\alpha}$ with photon index $\alpha \approx 2$, the measured count rate corresponds to an X-ray flux of $f_x = (7.1 \pm 0.2) \times 10^{-14}$ erg s$^{-1}$ cm$^{-2}$ within the 0.1–2.4 keV energy range. For this calculation, we used a column density of $N_H = 2 \times 10^{21}$ cm$^{-2}$ as inferred from the dispersion measure of $DM = 58.36$ pc cm$^{-3}$ (Lorimer et al. 1995) for a mean electron density of $\bar{n}_e = 0.03$ cm$^{-3}$ along the line of sight. The X-ray flux corresponds to a luminosity of $L_x = (5.3 \pm 0.2) \times 10^{31}$ erg/s assuming a pulsar distance of $d = 2.5$ kpc.

### *2.1. Neutron star surface temperature upper limit*

In the absence of a detailed spectrum, an upper limit to the neutron star surface temperature $T_s$ can be calculated from the observed PSPC count rate by assuming blackbody emission from the whole surface of the neutron star of radius $R$, distance $d$ and absorption column density $N_H$ (e.g. Becker, Trümper & Ögelman 1993b). Due to the gravitational redshift $z$, however, only $T_s^\infty = T_s/(1+z)$ can be observed. Theoretical calculations of cooling rates, i.e., the rate of thermal emission from the surface of a neutron star of given age, depend mainly on the internal structure of the star, in particular on the equation of state at super-nuclear densities. The possibility of altered hadronic interactions, the existence of stable pions, kaons, hyperons or free quarks in the core of a neutron star and the influence of superfluid neutrons in the inner crust all yield different cooling rates and theoretical predictions, so that the neutron star surface temperature as a function of the neutron star age reflects the stellar composition. Comparing the neutron star temperatures and temperature upper limits found by ROSAT with the theoretical predictions based on different equations of state thus provides the empirical basis essential for the verification of neutron star models and cooling theories. In the literature have been proposed. To make our results comparable with the predictions of these models, the stellar parameters $M$ (mass) and $R$ (radius) from these models have been used for computing $T_s^\infty$ and $L_\gamma^\infty$, respectively. Table 1 shows the results derived for the different cooling models which are developed, for example, in Nomoto & Tsuruta (1987), Van Riper (1991), Umeda et al (1993), Chong & Cheng (1993) and Page (1994).

**Table 1.** Surface temperature upper limits for PSR 2334+61

| Model | Radius km | Mass $M_\odot$ | $\log T_s^\infty$ K | $\log L_\gamma^\infty$ erg/s |
|---|---|---|---|---|
| PS[1] | 16.10 | 1.31 | $5.89^{+0.15}_{-0.10}$ | $32.95^{+0.60}_{-0.38}$ |
| PS[2] | 15.83 | 1.40 | $5.89^{+0.15}_{-0.10}$ | $32.95^{+0.60}_{-0.38}$ |
| MPA[3,4] | 12.45 | 1.40 | $5.91^{+0.15}_{-0.09}$ | $32.86^{+0.59}_{-0.36}$ |
| PAL33[4] | 11.91 | 1.40 | $5.91^{+0.15}_{-0.09}$ | $32.84^{+0.59}_{-0.36}$ |
| UV14[4,5] | 11.20 | 1.40 | $5.92^{+0.15}_{-0.09}$ | $32.82^{+0.59}_{-0.35}$ |
| UU[6] | 11.14 | 1.40 | $5.92^{+0.15}_{-0.09}$ | $32.82^{+0.59}_{-0.35}$ |
| PAL32[4] | 11.02 | 1.40 | $5.92^{+0.15}_{-0.09}$ | $32.82^{+0.59}_{-0.35}$ |
| FP[1] | 10.90 | 1.29 | $5.92^{+0.15}_{-0.09}$ | $32.81^{+0.59}_{-0.35}$ |
| FP[4,5] | 10.85 | 1.40 | $5.92^{+0.15}_{-0.09}$ | $32.80^{+0.59}_{-0.35}$ |
| AV14[4,5] | 10.60 | 1.40 | $5.92^{+0.15}_{-0.09}$ | $32.81^{+0.59}_{-0.35}$ |
| AU[6] | 10.40 | 1.40 | $5.92^{+0.15}_{-0.09}$ | $32.80^{+0.59}_{-0.35}$ |
| FP(pion)[1] | 9.40 | 1.27 | $5.93^{+0.15}_{-0.08}$ | $32.75^{+0.58}_{-0.34}$ |
| PAL(kaon)[7] | 8.20 | 1.40 | $5.94^{+0.15}_{-0.08}$ | $32.73^{+0.58}_{-0.33}$ |
| BPS[1] | 7.90 | 1.35 | $5.94^{+0.15}_{-0.08}$ | $32.72^{+0.58}_{-0.33}$ |
| BPS[2] | 7.35 | 1.40 | $5.94^{+0.15}_{-0.08}$ | $32.71^{+0.58}_{-0.33}$ |
| PAL(kaon)[7] | 7.00 | 1.40 | $5.94^{+0.15}_{-0.08}$ | $32.71^{+0.58}_{-0.33}$ |
| Polar Cap | 0.20 | 1.40 | $6.45^{+0.15}_{-0.07}$ | $31.48^{+0.58}_{-0.27}$ |

The stellar parameters radius and gravitational mass are taken from: [1]Umeda et al. (1993), [2]Van Riper (1991), [3]Müther et al. (1987), [4]Page (1994), [5]Wiringa et al. (1988), [6]Chong & Cheng (1993), [7]Thorsson et al. (1994). The allowed range quoted for $T$ and the bolometric luminosity $L_\gamma$ includes the uncertainties of distance and column density.

Using the dispersion measure based distance as lower limit and the largest SNR distance as upper limit, combined with the full range of column densities inferred from the dispersion measure and the galactic HI-model of Dickey & Lockman (1990); i.e. $d = 2.5 \pm 0.5$ kpc and $N_H = 2^{+6.9}_{-1.0} \times 10^{21}$ cm$^{-2}$, implies temperature upper limits which are consistent with the predictions of standard cooling models. None of the various neutron star models is excluded by these upper limits even if heating processes like frictional heating or crust cracking are taken into account. A comparison of the upper limits with cool-dard cooling and frictional heating is shown in Figure 1. The model shows agreement with the observation but only if the uncertainty in the distance and in the column density is stretched to the limits. Strong frictional heating in the PS model is excluded by our observation. This remains true even if we take into account that the pulsar's characteristic age ($P/2\dot{P} \approx 41,000$ year) may underestimate the true age by up to 20% due to the unmeasured braking index. A more detailed comparison between the different thermal evolution models and the temperature upper limits for the ROSAT detected rotation-powered pulsars can be found in Becker (1995).

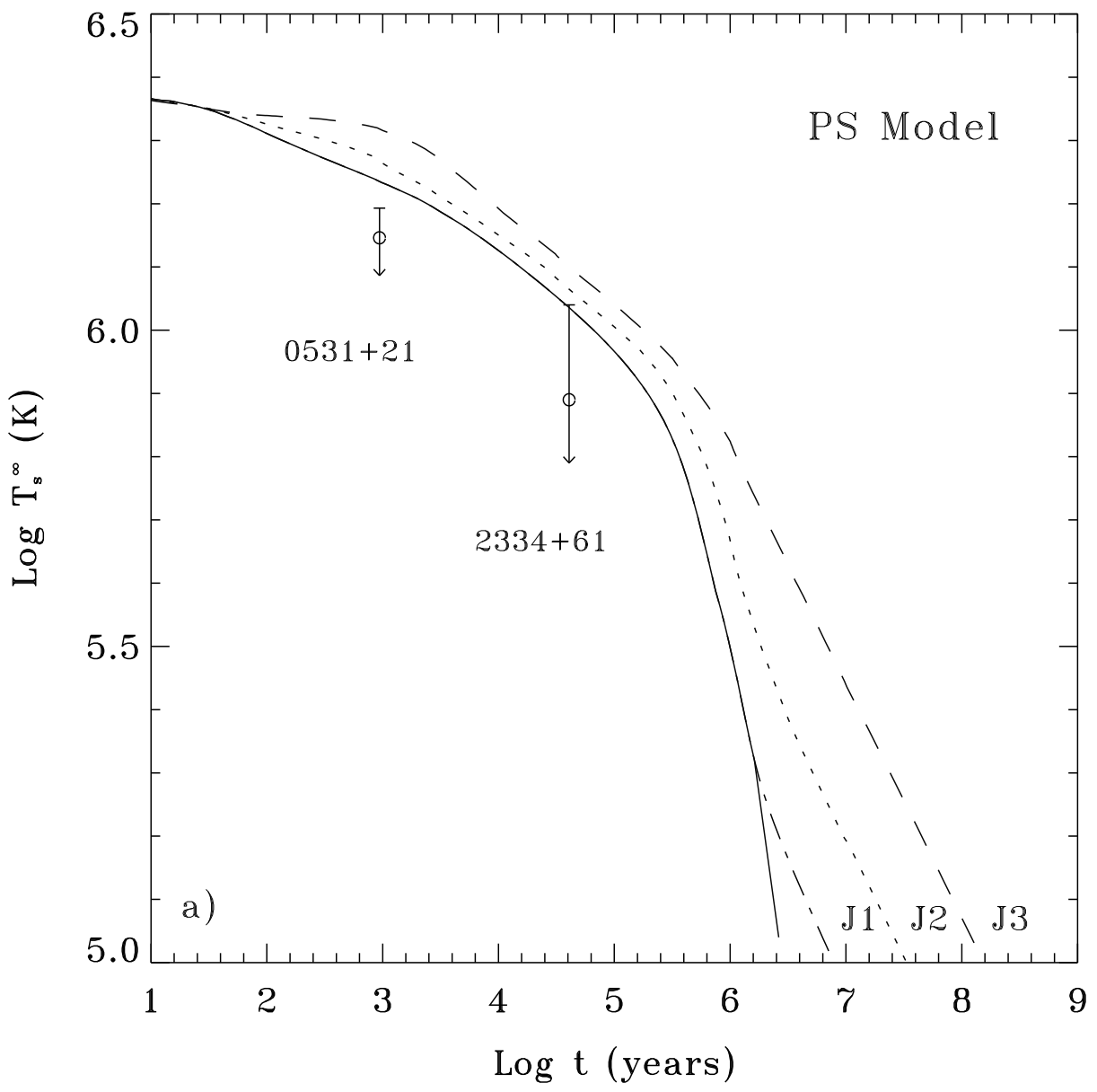

**Fig. 1.** Temperature upper limit for PSR 2334+61, using the PS model parameters from Umeda et al. (1993). Also shown are the cooling curves of this model and the temperature upper limit for the Crab pulsar (Becker & Aschenbach 1995). The upper limits are represented by a circle and the arrow lengths display the uncertainty range due to the uncertainties in distance and column density. The dotted and dashed lines show the predicted thermal evolution of the neutron star with internal frictional heating in the case of strong (J3), weak (J2) and superweak (J1) pinning of crustal superfluid vortex lines.

## 3. Conclusions

X-rays from PSR 2334+61 have been detected for the first time with the ROSAT PSPC detector. The source count rate of $(0.0018 \pm 0.0005)$ cts/s and an effective observation time of 8350 seconds were not sufficient for detailed analysis. Nevertheless, some estimates are possible if we make some assumptions. A thermal model, in which we assume that the X-ray flux is from the hot neutron star surface, gives a surface temperature in good agreement with the predictions of popular neutron star cooling mod-of emission from the pulsar magnetosphere or from a synchrotron nebula, gives a X-ray luminosity that is a very plausible fraction of the available spin-down power. Only a much deeper observation of this pulsar will distinguish between these options.

## Acknowledgments

The ROSAT project is supported by the BMFT. KTSB acknowledges receipt of a PPARC postdoctoral research fellowship. The data were analyzed using the software package EXSAS.

## References

Becker, W., Trümper, J. & Ögelman, H., 1993a, IAU Circular No. 5805

Becker, W., Trümper, J. & Ögelman, H., 1993b, in *Isolated Pulsars*, eds K.A. Van Riper, R.I. Epstein & C. Ho, 104, Cambridge University Press

Becker, W., Aschenbach, B., 1995, in *The Lives of Neutron Stars*, eds A. Alpar, U. Kilizóglu & J. van Paradijs, Kluwer Academic Publishers

Becker, W., 1995, PhD thesis, Ludwig-Maximilians-Universität München, available as MPE-Report 260

Becker, W., Brazier, K.T.S. & Trümper, J., 1995, A&A, **298**, 528

Dickey, J. M., and Lockman, F. J. 1990, Ann. Rev. Astron. Astrophys., **28**, 215

Kulkarni, S., Predehl, P.R., Hasinger, G. & Aschenbach, B.R., 1993, Nat, **362**, 135

Lorimer, D.R., Yates, J.A., Lyne, A.G. & Goukd, D.M., 1995, MNRAS, **273**, 411

Reich, W., Braunsfurth, E., 1981, A&A, **99**, 17

Fürst, E., Reich, W. & Seiradakis, J.H (1993), A&A, **276**, 470

Müther, H.,Prakash, M. & Ainsworth, T.L., 1987, Phys. Lett. B119, 469

Nomoto, K. & Tsuruta, S., 1987, ApJ, 312, 711

Page, D., 1994, ApJ (7/94)

Ögelman, H., Finley, J.P. & Zimmermann, H.U., 1993, Nat, **361**, 136

Ögelman, H.B., 1995, in *The Lives of Neutron Stars*, eds A. Alpar, U. Kilizóglu & J. van Paradijs, Kluwer Academic Publishers

Taylor, J.H., Cordes, J.M., 1993, ApJ, **411** 674

Taylor, J., Lyne, A.G. & Manchester, R.M., 1995, Princeton-Pulsar-catalogue

Thorsson, Prakash, M. & Lattimer, J.M., 1994, preprint

Umeda, H., Shibazaki, N., Nomoto, K. & Tsuruta, S., 1993, ApJ, 408, 286

Van Riper, K.A., 1991, ApJS, 74, 449

Wiringa, R.B., Fiks, V. & Fabrocini, A., 1988, Phys. Rev. C38, 1010

# ROSAT observations of PSR 2334+61 in the supernova remnant G114.3+0.3

W. Becker[1], K.T.S. Brazier[2] & J. Trümper[1]

[1] Max-Planck-Institut für extraterrestrische Physik
85740 Garching bei München, Germany

[2] Institute for Astronomy, University of Edinburgh,
Blackford Hill, Edinburg, EH9 2HJ